# Relative Distribution of Water Clusters at Temperature (300-3000K) and Pressure (1-500MPa)

Ri Yong-U, Kwak Chang-Sob, Sin Kye-Ryong [*]

( *Faculty of Chemistry, **Kim Il Sung** University, Pyongyang, DPR of Korea* )

* ryongnam9@yahoo.com

**Abstract**: At 300-3000K and 1-500MPa, variations of relative contents for small water clusters $(H_2O)_n$ ( n=1~6 ) were calculated by using statistical mechanical methods. First, 9 kinds of small water clusters were selected and their structures were optimized by using *ab initio* method. In the wide range of temperature (300-3000K) and pressure (1-500MPa), their equilibrium constants of reactions for formation of 9 kinds of water clusters were determined by using molecular partition function. Next, changes of contents (molar fractions) as function of temperature and pressure were estimated. The obtained results for small water clusters can be used to interpret temperature-pressure dependency of the average number for the hydrogen bonds in water clusters and redistribution of the water clusters at the ultrasonic cavitation reactions.

## 1. Introduction

For some decades, there have been many efforts for transforming micro-structures of liquid water clusters by processing with several experimental methods[1,3] ( i.e. magnetic, ultrasonic, hydrodynamic method and et al ) and for discovering the reaction mechanisms between water clusters at different conditions by using the theoretical methods[4,5,6,9].

When liquid water is treated ultrasonically, the formation and explosion of numerous cavitation bubbles subject liquid water to intense heat and pressure. This was regarded as one of the reasons that micro-structures of liquid water clusters were transformed, but the detailed researches on their mechanisms were rarely presented[7,8,10].

In this paper, it was estimated and analyzed the changes of relative contents of small water clusters generated under 300-3000K and 1-500MPa by using ab initio method and statisti cal mechanical method.

## 1. Models and Computations
### 2.1 Calculation of equilibrium constants

Here models for the water clusters $(H_2O)_n$ (n=1-6) were composed of 9 small water clusters including water monomer and their geometric structures were selected from the results of the global potentials minimization (Fig. 1)[11,12] by ab initio (HF/6-31G) method in the quantum chemical software Gaussian03[2].

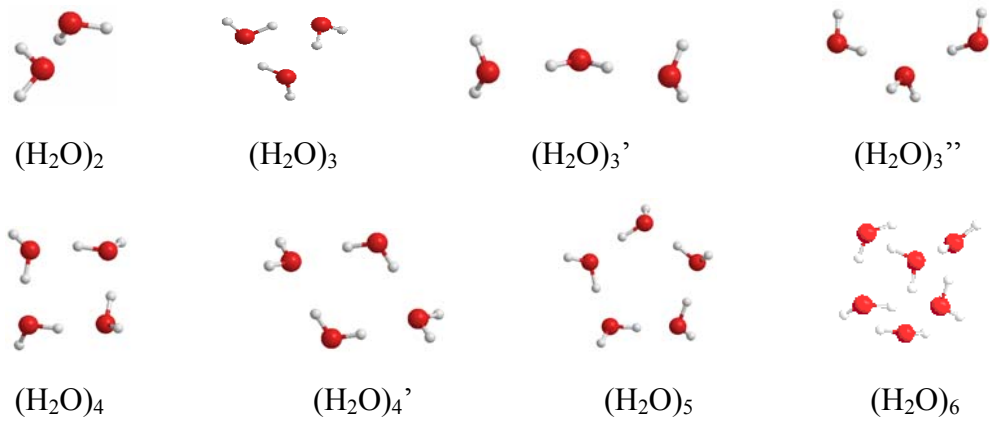

(H$_2$O)$_2$    (H$_2$O)$_3$    (H$_2$O)$_3$'    (H$_2$O)$_3$''

(H$_2$O)$_4$    (H$_2$O)$_4$'    (H$_2$O)$_5$    (H$_2$O)$_6$

Fig. 1    Different models for small water clusters

When the partition function was calculated, each water cluster was assumed to form ideal gas and harmonic oscillator and rigid rotator approach was used.

At the ground state, the electronic energy of a molecule was taken to be zero and if there were not degeneracy for the electronic energy (in case of water clusters, total spin S=0), the value of electron partition function was equal to 1.

The rotational partition function ($q_r$) was calculated as follows:

$$q_r = \frac{\sqrt{\pi}}{\sigma}(\frac{8\pi kT}{h^2})^{\frac{3}{2}}\sqrt{I_x I_y I_z} \qquad (1)$$

where k is Boltzmann constant and $h$ is Planck's constant, and principal moments of inertia ($I_x$, $I_y$, $I_z$) were calculated based on the optimized structures of the water clusters. Symmetry number ($\sigma$) derives from their structural symmetry.

The vibrational partition function ($q_v$) was calculated as follows:

$$q_V = \prod_{i=1}^{3n-6} \frac{\exp(-h\nu_i/2kT)}{1-\exp(-h\nu_i/kT)} \qquad (2)$$

where n is the number of atoms in a water cluster and $\nu_i$ is the normal vibration frequency. $\nu_i$ was determined form the calculated vibrational spectrum by using HF/6-31G.

Statistical mechanical simulations were carried out in the reaction system containing 9 kinds of the small water clusters in Fig. 1.

Here it was assumed that water clusters (n>1) were created by bimolecular reactions between two kinds of water clusters that have similar geometric structures and seem to have lower activation energy. Among many possible bimolecular reactions for generating the water clusters, chosen were 8 reactions which have larger equilibrium constant (see Table 1). These reactions were used for calculating the equillibrium distributions of the water clusters as shown below.

Table 1. Selected bimolecular reactions and their equilibrium constants

| № | Reactions | Equilibrium constants |
|---|---|---|
| 1 | $H_2O + H_2O = (H_2O)_2$ | $K_1 = x_2/x_1^2$ |
| 2 | $(H_2O)_2 + H_2O = (H_2O)_3$ | $K_2 = x_3/x_1 x_2$ |
| 3 | $(H_2O)_2 + H_2O = (H_2O)_3'$ | $K_3 = x_3'/x_1 x_2$ |
| 4 | $(H_2O)_2 + H_2O = (H_2O)_3''$ | $K_4 = x_3''/x_1 x_2$ |
| 5 | $(H_2O)_2 + (H_2O)_2 = (H_2O)_4$ | $K_5 = x_4/x_2^2$ |
| 6 | $(H_2O)_2 + (H_2O)_2 = (H_2O)_4'$ | $K_6 = x_4'/x_2^2$ |
| 7 | $(H_2O)_3 + (H_2O)_2 = (H_2O)_5$ | $K_7 = x_5/x_2 x_3$ |
| 8 | $(H_2O)_3 + (H_2O)_3 = (H_2O)_6$ | $K_8 = x_6/x_3^2$ |

( $x_i$ : molar fraction of i-th cluster)

Equiliblium constant ($K^{\ominus}$) of the reactions was calculated as follows:

$$\ln K^{\ominus} = -\frac{\Delta \varepsilon_0}{kT} + \ln \prod_{\alpha}(\frac{kT}{P^{\ominus} \Lambda_{\alpha}^3} q_{int}^{\alpha})^{\nu_{\alpha}} \qquad (3)$$

$$\Lambda = (h^2/2\pi mkT)^{1/2}$$

where $P^{\ominus}$ is the standard pressure ($10^5$ Pa), $\Lambda$ is the thermal wavelength, α notates chemical species and $\nu_{\alpha}$ means the stoichiometric coefficient of α species at reaction ( negative value for reactant and positive value for product). $q_{int}^{\alpha}$ is the internal partition function of molecule and calculated as product of electronic, vibrational, and rotational partition functions. $\Delta \varepsilon_0 = \sum_{\alpha} \nu_{\alpha} \varepsilon_0^{\alpha}$, where $\varepsilon_0^{\alpha}$ is the electronic energy of molecule at the ground state.

**2. 2 Calculation of equilibrium distributions of water clusters under various temperatures and pressures**

From the meaning of $x_i$ in Table 1, it is easy to obtain Eq. 4 as follows:

$$x_1 + x_2 + x_3 + x_3' + x_3'' + x_4 + x_4' + x_5 + x_6 = 1 \tag{4}$$

For determination of the equilibrium distributions of water clusters, it needs to find out the linear independent reactions among the possible bimolecular reactions between water clusters, the components of the reaction system. Here, by calculating the ranks of the reaction matrix, 8 linear independent reactions were chosen.

Equations for the equilibrium constants of the 8 linear independent reactions can be expressed by using $K_i$ (i=1~8) in Table 1 and combined with Eq. 4 into Eq. 5 as follows:

$$K_1^2 K_2^2 K_8 x_1^6 + K_1^2 K_2^2 K_7 x_1^5 + (K_1^2 K_5 + K_1^2 K_6) x_1^4 + \\ + (K_1 K_2 + K_1 K_3 + K_1 K_4) x_1^3 + K_1 x_1^2 + x_1 - 1 = 0 \tag{5}$$

By using $K_i$ (i=1~8) calculated with the molecular partition functions, Eq. 5 was solved numerically to give 6 answers for $x_1$, that is, 4 complex numbers, one positive (between 0 and 1), and one negative number, respectively. Here the positive number is molar fraction of water monomer, $H_2O$. Molar fractions of other water clusters in Fig. 1 can be also calculated by using the similar procedure.

As the equilibrium constants ( $K_i$ in Table 1 ) are expressed with molar fractions, it is necessary to express them with pressure like Eq. 6 as follows:

$$K_i = K^{\ominus} P / P^{\ominus} \tag{6}$$

or

$$K_i = K^{\ominus} (P/P^{\ominus}) \exp(-P\Delta B / RT) \tag{7}$$

$$\Delta B = 4 N_A \Delta V$$

where $P^{\ominus}$ is standard pressure (0.1MPa) and $P$ is the total pressure of reaction system. $\Delta B$ is the difference of the second virial coefficients of product and reactant molecule. $\Delta V$ is difference of the volumes of product and reactant molecule and $N_A$ is Avogadro's constant.

Eq. 6 estimates the pressure of the system under the ideal gas approach and Eq. 7 estimates it by fugacity value of the system under the hard-sphere potential approach.

By using Eq. 6 and 7, calculated were the equilibrium constants at the different temperatures and pressures, and followed by the calculations for variations of molar fractions of water clusters.

## 3. Results and Discusion

### 3.1 Relative distribution of small water clusters on the temperatures and pressures

Distribution of the water clusters calculated under different pressures by Eq. 6 and 7 were not different in lower pressures (0.1-10³ MPa), but quite different under higher pressure. Fig. 2 and 3 show the calculated results under the pressure of $10^4$ MPa.

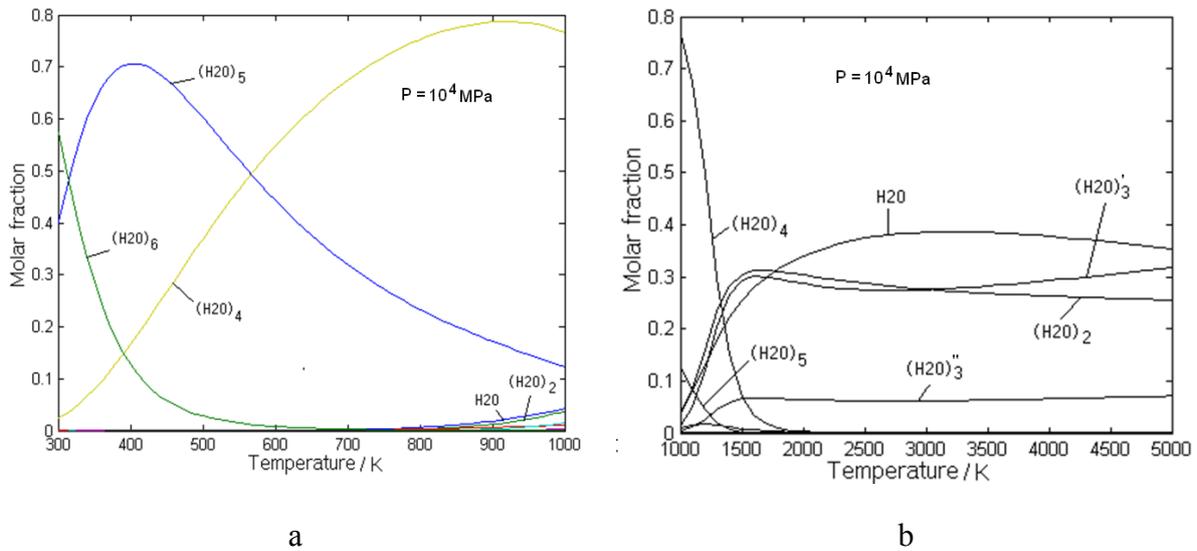

a b

Fig. 2 Relative distribution of water clusters under the ideal gas approach
(a: 300~1000K, b: 1000~5000K)

Under the temperature above 1500K, tetramers, pentamers, and hexamers disappear and monomers, dimers, and trimers can exist and the content ratio of tetramers to pentamers is different below 1500K.

To discuss ultrasonic cavitation effects, the equilibrium compositions were calculated at wide range of pressure (1~500MPa) and temperature (300~5000K) by using Eq. 7. Under these temperatures and pressures, content of trimers was very small and the contents of monomer, dimer, tetramer, pentamer and hexamer were remarkably changed. Fig. 4 and 5 showed change of contents of $H_2O$ and $(H_2O)_2$ respectively and Fig. 6 showed those for 4-, 5- and 6-mers.

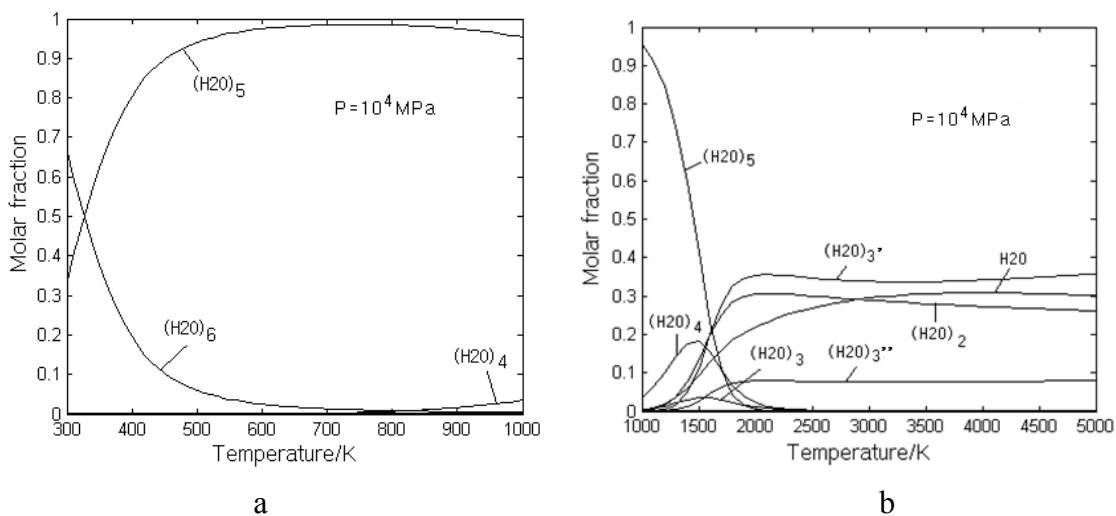

a b

Fig. 3  Relative distribution of water clusters accounted with fugacity.
(a: 300~1000K, b: 1000~5000K)

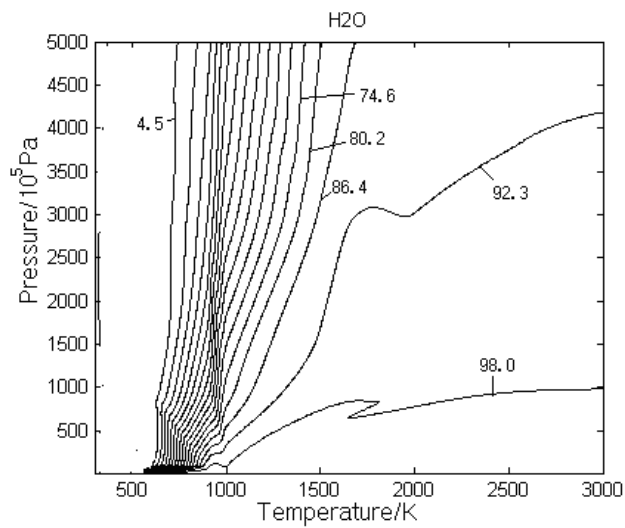

Fig. 4  Temperature-pressure dependency of content (%) of $H_2O$

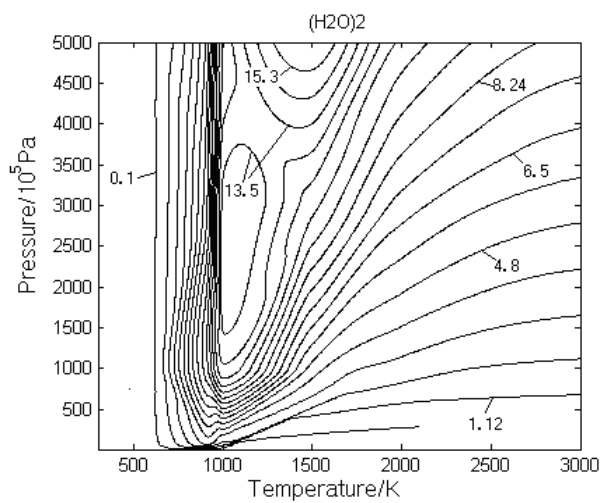

Fig. 5  Temperature-pressure dependency of content (%) of $(H_2O)_2$

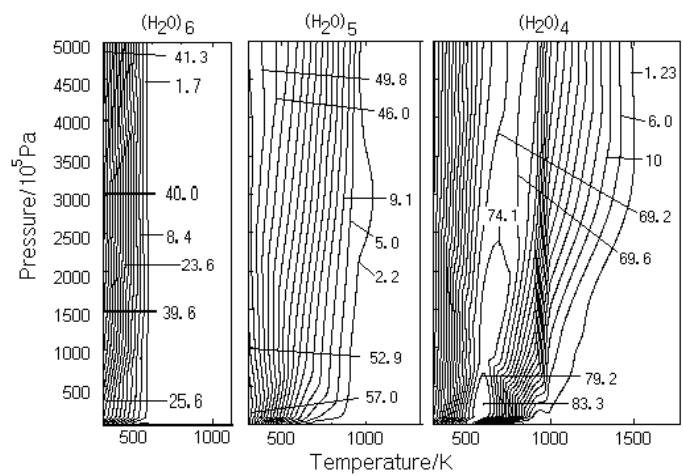

Fig. 6  Temperature-pressure dependency of contents (%) of $(H_2O)_4$, $(H_2O)_5$ and $(H_2O)_6$

## 3.2 Relationship between the number of hydrogen bonds and the relative distribution of water clusters

The calculated values of the average number of the hydrogen bonds and the relative distribution of the water clusters were compared with those from the experimental data in Table 2[13].

Table 2. The average hydrogen bond number of each water cluster ($n_H$)

| № | Cluster | $n_H$ |
|---|---|---|
| 1 | $H_2O$ | 0 |
| 2 | $(H_2O)_2$ | 1 |
| 3 | $(H_2O)_3$ | 2 |
| 4 | $(H_2O)_{3'}$ | 1.33 |
| 5 | $(H_2O)_{3''}$ | 1.33 |
| 6 | $(H_2O)_4$ | 2 |
| 7 | $(H_2O)_{4'}$ | 2 |
| 8 | $(H_2O)_5$ | 2 |
| 9 | $(H_2O)_6$ (prism type) | 3 |

If molar fraction of each cluster is given, the average hydrogen bond number of the water clusters system can be calculated as follows:

$$N_H = \sum_{i=1}^{9} x_i \cdot n_H(i) \qquad (8)$$

where $x_i$ is molar fraction and $n_H(i)$ is the average hydrogen bond number of the i-th cluster.

$N_H$ calculated in the temperature range (0~100℃) were compared with the experimental ones[13] in Fig. 7, where the calculated ones were larger than the experiment, but the temperature dependency of $N_H$ was quite similar.

It means that the present method for the water cluster system can be applied in interpretation for the behavior of the hydrogen bond system under high temperature and high pressure such as the ultrasonic cavitation effects.

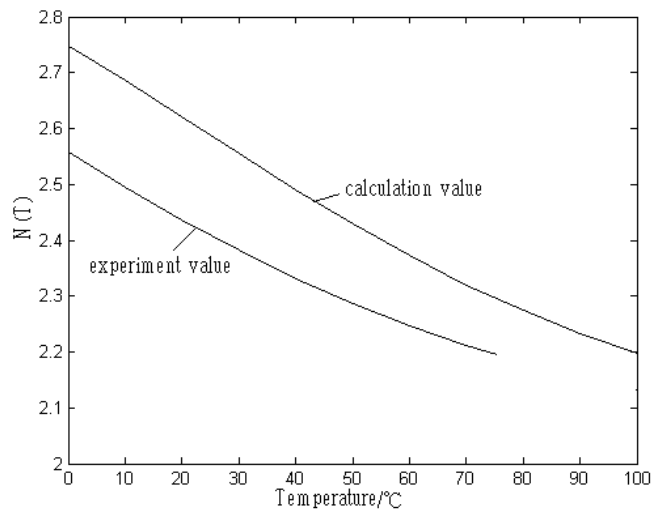

Fig. 7  Comparison with the mean hydrogen bond numbers from the calculated and experiment data

## 4. Conclusion

For the hydrogen bond system with the small water clusters $(H_2O)_n$ (n=1-6), the relative contents of 9 stable water clusters and their temperature-pressure dependency were calculated by using molecular partition function. If the temperature was increased, the big water clusters were broken first and the contents of dimer and trimer were gradually increased. The present method can be used in interpreting temperature-pressure dependency of the average hydrogen-bond number and the redistribution of water clusters at the ultrasonic cavitation reactions.